\documentclass[11pt,showpacs,showkeys,nofootinbib,preprintnumbers]{revtex4-2}
\usepackage{graphicx}   
\usepackage{color}         
\usepackage{braket}
\usepackage{amsmath}
\usepackage{amssymb}
\usepackage{mathrsfs} 
\usepackage{titlesec}
\usepackage{makecell}
\usepackage{natbib,hyperref}
\usepackage[mathscr]{euscript}
\usepackage{dutchcal}
\usepackage{mathrsfs}  
\usepackage{bm} 
\usepackage[T1]{fontenc}
\usepackage[utf8]{inputenc}
\usepackage[english,german]{babel}

\def \bea {\begin{eqnarray}}
\def \eea {\end{eqnarray}}

\newcommand{\R}{\mathbb{R}}
\newcommand{\nn}{\nonumber}

\definecolor{ao-english}{rgb}{0.0, 0.5, 0.0}
\definecolor{cadmiumblue}{rgb}{0.0, 0.42, 0.24}

\begin{document}

\preprint{ECTP-2026-11}
\preprint{WLCAPP-2026-11}
\hspace{0.05cm}

\title{Stress-Energy Tensor for Modified General Relativity with Quantum-Deformed Metric in Riemann Spacetime}
\email{Corresponding author: tawfik@itp.uni-frankfurt.de; 400778@iu.edu.sa; atawfik@bnl.gov}

\author{A. Tawfik$^{1,2}$, Saleh O. Allehabi$^{1}$, A. A. Alshehri$^{3}$, M. Nasar$^{4}$, M. Maher$^{5}$}

\affiliation{$^1$Department of Physics, Faculty of Science, Islamic University of Madinah, Madinah 42351, Saudi Arabia} 
\affiliation{$^2$Basic Science Department, Faculty of Engineering, Ahram Canadian University (ACU), 12556 Giza, Egypt}
\affiliation{$^3$Department of Science and Technology, University College at Nairiyah, University of Hafr Al Batin (UHB), Nairiyah 31981, Saudi Arabia} 
\affiliation{$^4$Physics Department, Faculty of Science, Benha University, 13511 Benha, Egypy} 
\affiliation{$^{5}$Physics Department, Faculty of Science, Capital University, 11792 Ain Helwan, Egypt} 

\begin{abstract}
The generalized noncommutative Heisenberg algebra which is based on the generalized uncertainty principle, imposes a minimal length uncertainty on quantum mechanics (QM), on one hand. On the other hand, the quantum-induced spacetime which is based on quantum-deformed metric through geometric quantization is proposed as additional curvatures on the relativistic tangent bundle on Finsler manifold. An additional terms that incorporates minimal length discretization along with second-order derivatives of tangent covectors, thereby suggesting an approach to reconcile the principles of QM with General Relativity (GR), is utilized for the construction of a torsion-free quantum-deformed metric on Riemann manifold. Consequently, it is proposed that quantum-induced revisions to the symmetric stress-energy tensor, source of spacetime curvature, along with the current density related to the gauge transformations of gravity, ought to be taken into account in the matter Lagrangian with electromagnetic and scalar components. Vanishing covariant derivative of the quantum-induced stress-energy tensor suggests that the corresponding continuity equation implies that the gravitational fields do work on the classical and quantum matter and vice versa and the non-gravitational energy and momentum are no longer entirely conserved. For vanishing tangent covector derivatives and/or minimal length uncertainty, the classical formulations of the Einstein stress-energy tensor are retained and accordingly that of GR and QM. We conclude that the proposed quantum-induced formulation of the stress-energy tensor is fundamentally suitable for both classical and quantum-induced field equations.
\end{abstract}

\date{\today}


\maketitle


\section{Introduction} 
\label{sec:intro}

While Special Relativity (SR) and Quantum Mechanics (QM) were reconciled long ago \cite{10.1098/rspa.1930.0013}, there has been no successful attempt in the past century to unify General Relativity (GR) with QM \cite{Mohageg2025}. The present manuscript relies on approaches established by Born  \cite{Born:1935:Nature,born1938suggestion,Born:1949dbq,Born:1949yva} in the 1930s and 1940s, along with those by Caianiello \cite{Caianiello:1980iv,Caianiello:1981jq,caianiello1984maximal,Caianiello:1989wm} in the 1980s of the last century. We also utilize the recent progresses in quantum geometry \cite{Rosen:1962mpv,Brandt:1991hy,Brandt:1991sw,Ashtekar:1998ak,Brandt:2000gka} and noncommutative algebra \cite{Mead:1966zz,Snyder:1946qz,Majid:1990gq,Majid:1999tc,Majid:1994cy,Kempf:1994su,Maggiore:1993zu,Maggiore:1993rv}. For a source of spacetime curvature and current density associated with the gauge transformations of gravity, we assume symmetric stress-energy tensor. With the quantum-induced torsion-free metric tensor recently introduced \cite{Tawfik:2025wsl,Tawfik:2025ldp,Tawfik:2024itt,Tawfik:2024bdf,NasserTawfik:2024afw,Farouk:2023hmi,Tawfik:2023ugm,Tawfik:2023hdi,Tawfik:2023orl,Tawfik:2023kxq,Tawfik:2020zvf,TawfikAN2023b,
TawfikAN2023a}, a geometric quantization ansatz, the Einstein field equations can be constructed. 

The geometric quantization, which focuses on the fundamental metric as its primary target for quantization, offers a promising pathway toward reconstructing GR from a quantum-induced perspective. This innovative ansatz utilizes the intrinsic geometric properties of spacetime \cite{Ambjorn:1997vf,Ketels:2025uot}, enabling a more direct and potentially more accurate representation of GR phenomena at the quantum level \cite{Nakanishi:1977gt,Boulware:1974sr,Osborne:1949zz}. One of the significant advantages of this technique is its applicability in regimes where conventional formulations of GR encounter limitations or unresolved challenges, such as at singularities or in high-energy limits where classical theories break down \cite{Tawfik:2025icy,Tawfik:2025rel,Tawfik:2025kae,Tawfik:2023onh}. Unlike other alternative approaches, such as loop quantum gravity, which rely on the discretization of space or other complex quantum structures \cite{Rovelli:1997yv,Meissner:2004ju}, or modified theories of gravity that introduce additional fields or parameters \cite{Chokyi:2026ymi,Ghosh:2026rll,Chokyi:2024wqz}, the geometric quantization technique maintains a broad and highly adaptable scale. Its wide range of applicability arises from its profound reliance on the metric tensor, which is central to the geometric description of spacetime, making it a very generic and adaptable framework. Consequently, it can potentially be utilized across a variety of scenarios - from cosmological models to black hole physics - without necessitating extensive modifications or assumptions. This universality not only boosts its theoretical attractiveness but also paves the way for investigating quantum gravitational effects in diverse physical contexts, offering insights that could unifies the classical and quantum descriptions of gravity in a coherent and comprehensive manner \cite{Eichhorn:2019dhg}. This study examines the impact of the resulting quantum-deformed metric on the stress-energy tensor, considering both matter and electromagnetic fields. Some observable gravitational effects and potential phenomenological applications are proposed. In other publications, we have derived alterations in all quantities or facets of general relativity, including affine connections and curvatures, i.e., Einstein tensor \cite{Tawfik:2025wsl,Tawfik:2025ldp,Tawfik:2024itt,Tawfik:2024bdf,NasserTawfik:2024afw,Farouk:2023hmi,Tawfik:2023ugm,Tawfik:2023hdi,Tawfik:2023orl,Tawfik:2023kxq,Tawfik:2020zvf,TawfikAN2023b,
TawfikAN2023a}.

Under the specified conditions for geometric quantization, the continuity equation, which signifies the local conservation of energy and momentum, assumes a modified form that illustrates the interactions between gravitational fields and matter - both classical and quantum \cite{Bagchi:2001qu}. This implies that gravitational fields can perform work on matter, thereby affecting its energy and momentum, while matter can also modify the gravitational field through its energy-momentum content \cite{DeWitt:1952bs}. Consequently, the traditional notion of absolute conservation of non-gravitational energy and momentum is invalidated in this quantum-gravitational context. Instead, the interchange of energy and momentum between gravitational fields and matter introduces a dynamic relationship, resulting in a situation where the total non-gravitational energy and momentum are no longer strictly conserved locally \cite{Westpfahl:1987hwd}. This transformation has profound implications for our understanding of energy conservation laws in a quantum gravitational setting and suggests that the interplay between matter and spacetime geometry is more complex and less absolute than in classical theories \cite{Komar:1958wp,Goldberg:1958zz}.

This paper presents a uncompounded formulation of the quantum-deformed metric that is directly related to the concept of minimal length deduced from generalized uncertainty principle \cite{Tawfik:2024gxc}. Although it may look simple in relation to the formulations currently utilized
\cite{Tawfik:2025wsl,Tawfik:2025ldp,Tawfik:2024itt,Tawfik:2024bdf,NasserTawfik:2024afw,Farouk:2023hmi,Tawfik:2023ugm,Tawfik:2023hdi,Tawfik:2023orl,Tawfik:2023kxq,Tawfik:2020zvf}, the physical motivation for both formulations is common, an approximate determination of the conformal coefficient. The present manuscript is organized as follows: The formalism is outlined in section \ref{sec:gupAffine}. Section \ref{sec:rgup} is devoted to the relativistic generalized uncertainty principle in Finsler geometry \cite{Chen1996}.
The quantum-induced  stress-energy tensor is introduced in section \ref{sec:tmunu1}. The quantum-induced corrections to the matter Lagrangian are given in section \ref{sec:qLagr}. Section \ref{sec:qLagrEM} elaborates the quantum-induced corrections to the stress-energy tensor with an electromagnetic Lagrangian. Symmetry properties and covariant derivative of stress-energy tensor with an electromagnetic Lagrangian are outlined in section \ref{sec:symmTEM} and \ref{sec:cdTEM}, respectively. The quantum-induced  stress-energy tensor with a scalar Lagrangian and its  symmetry property and covariant derivative is worked out in section \ref{sec:QMscal}, \ref{sec:symmTscal}, and \ref{sec:codrTEM}, respectively. Section \ref{sec:Cncls} is devoted to the conclusions and outlook.

\section{Formalism}
\label{sec:gupAffine}

The formalism is initiated by a brief review of the Relativistic Generalized Uncertainty Principle (RGUP) \cite{Tawfik:2024gxc} in following section.

\subsection{Brief Review of Relativistic Generalized Uncertainty Principle}
\label{sec:rgup}

With isotropic property of spacetime and Snyder algebra \cite{Snyder:1946qz}, the relativistic generalized noncommutation relation is suggested as \cite{Tawfik:2024gxc,Todorinov:2020jtq,phdthesisXun}
\bea
\left[x^{\mu}, p^{\nu}\right] &=& i \hbar \left[\left(1 + \beta p_0^{\rho} p_{0 \rho}\right) \eta^{\mu \nu} + 2 \beta p_0^{\mu} p^{0 \nu}\right],
\eea 
where the subscript $0$ refers to auxiliary four-vectors. In four-dimensional flat spacetime, the geometry is defined by the Minkowski metric tensor $\eta^{\mu \nu}$. In the context of curved Riemann geometry, the relevant metric tensor is  $g_{\mu \nu}$. The transition from flat to curved spacetime can't be achieved merely by substituting $\eta_{\mu \nu}$ with $g_{\mu \nu}$; it necessitates the utilization of local inertial frames and the feasibility of coordinate transformations. 

In a given point within curved spacetime, one can exclusively adopt a coordinate system in which the metric tensor reduces to the Minkowski form, $\eta_{\mu \nu} = g_{\mu \nu}$, and its covariant derivatives vanish, $\nabla_{\alpha} g_{\mu \nu} = 0$. This local choice of coordinates, rooted in the equivalence principle of GR, underscores the fundamental notion that, at any specific point, the laws of physics can be expressed in the form of SR, reflecting the profound harmony between local inertial frames and the geometric fabric of spacetime. This implies that at any specific point $p$, it is possible to select a coordinate system - specifically, Riemann normal coordinates - in which the metric tensor simplifies to the Minkowski form $g_{\mu \nu} = \eta_{\mu \nu}$ and the Christoffel symbols $\Gamma^{\rho}_{\mu\nu}$ vanish. Conversely, under a general coordinate transformation $x^{\alpha} \rightarrow x^{\mu}(x)$, we find that
\bea
g_{\mu\nu}=\frac{\partial x^{\alpha}}{\partial x^{\mu}} \frac{\partial x^{\beta}}{\partial x^{\nu}} \eta_{\mu\nu}
\eea
Even though this formula seems to be globally applicable, in the scenario of curved spacetime, a single global transformation cannot be identified. A precise mathematical connection between curved and flat metrics may be represented by Taylor expansion,
\bea
g_{\mu\nu}(x)=\eta_{\mu\nu}-\frac{1}{3}R_{\mu\alpha\nu\beta}(p)x^{\alpha}x^{\beta}+{\cal O}(x^3).
\eea
The zeroth order is represented by $\eta_{\mu\nu}$, the first order is null, whereas the second order signifies the curvature. Additionally, a different general relationship is provided by tetrads,
\bea
g_{\mu\nu}(x)=e^{\alpha}_{\mu} e^{\beta}_{\nu} \eta_{\alpha\beta}.
\eea

In curved spacetime, the noncommutation relation between position and momentum, for example, establishes their uncertainties as derived from Robertson \cite{Rebertson1929} and the Schr{\"o}dinger uncertainty relation \cite{Schroedinger1930a},
\bea
\Delta x_0^{\mu}\, \Delta p_0^{\nu} &\gtrsim& \frac{\hbar}{2}\left[g^{\mu \nu} + \beta \left(\Delta p_0\right)^2 + \beta \langle p_0\rangle^2 - \beta \left(\Delta p_0^{\mu}\right)^2 - \beta \left(\Delta p_0^{\nu}\right)^2 \right]. \label{eq:2ordereq1}
\eea
To ensure that the quadratic equation,  Eq. (\ref{eq:2ordereq1}), possesses real roots for $\Delta p_0^{\nu}$,
\bea
\left(\Delta x_0^{\mu}\right)^2 &\gtrsim& \hbar^2 \left[\beta^2\left(\langle p_0\rangle^2 - \left(\Delta p_0^{\mu}\right)^2 -  \left(\Delta p_0^{\nu}\right)^2 \right) - \beta\, g^{\mu \nu}\right].
\eea
This inequality establishes a boundary for the minimal measurable length uncertainty. Moreover, when $\beta$, the RGUP parameter, is small, the higher-order terms become insignificantly small; thus, the uncertainty in the minimal measurable length can be approximated as
\bea
\Delta x^{\mu}_{\mathtt{min}} &\gtrsim& \pm \sqrt{-g^{\mu \nu}}\, \hbar \sqrt{\beta}. \label{eq:deltaxmin2}
\eea 
With $g^{\mu \nu}:=\left(g^{-1}\right)^{\mu \nu}$ is contravariant symmetric $(2,0)$-tensor field and $g:=\mathtt{det}\, g_{\mu \nu}$, one finds that
\bea
\Delta x^{\mu}_{\mathtt{min}} &\gtrsim& \pm \sqrt{-g}\, \sqrt{\beta_0}\, \ell_{\mathtt{p}}.
\eea 

Caianiello indicated that the quantization of GR in four-dimensional geometry may be feasible if one considers additional curvatures in the relativistic eight-dimensional spacetime tangent bundle $TM_8=M\otimes TM_4$ \cite{Caianiello:1980iv,Caianiello:1989wm,Caianiello:1989pu,Cainiello:1991htb}, where $TM_4$ refers to the four-dimensional tangent bundle. On the manifold $TM_8$, there are two features to be highlighted; (i) the covariant derivatives reflect the Lie algebra of the Heisenberg group, recognized as the Heisenberg algebra, which delineates the Heisenberg commutation relations in QM \cite{Hervik:2003vx}, and (ii) the quantum commutation relations are represented by the components of the curvature tensor \cite{Scarpetta2006}. 
For a free particle, the proposed $TM_8$ tangent bundle satisfies the duality-symmetry configuration of the spacetime coordinates and momenta, i.e., an eight-dimensional phase space, the Born reciprocity conjecture \cite{Born:1949dbq}.  An example of $TM_8$ would be Finsler manifold which is characterized by $M$ and the Finsler structure $F$, while Riemann manifold is characterized by $M$ and $g_{\mu \nu}$. At the local coordinates, $F$ is given as \cite{Chern1996,bao2000introduction}
\bea 
F^2(x, \dot{x}) &=& g_{\mu \nu}(x)\dot{x}^{\mu} \dot{x}^{\nu}, \label{eq:FStrctr1}
\eea 
where $x$ represent coordinates on tangent manifold and $\dot{x}$ their directions $\dot{x}=\partial x/\partial \zeta$ with $\zeta$ is a proper parametrization. The homogeneous property of $F$ in $\dot{x}$ allows for $F(x, \lambda\, \dot{x})=\lambda\, F(x, \dot{x}), \forall \lambda \in \R^+$ \cite{Capozziello:1999wx,bao2000introduction}. 
Accordingly, let us assume that 
\bea
F(x, (\sqrt{-|g|}\, \hbar \sqrt{\beta_0})\, \dot{x}) &=& (\sqrt{-|g|}\, \hbar \sqrt{\beta_0})\, F(x, \dot{x}), \qquad  \forall (\sqrt{-|g|}\, \hbar \sqrt{\beta_0}) \in \R^+ 
\eea

On $TM_8$, the metric tensor on tangent manifold $g_{ab} =g_{\mu \nu}\otimes g_{\mu \nu}$, where the basis coordinates read $x^a=(x^{\mu},\dot{x}^{\nu})$ could be determined by the Hessian matrix,
\bea
g_{ab}(x) &=& \frac{1}{2} \frac{\partial^2}{\partial \dot{x}^{a}\, \partial \dot{x}^{b}} F^2(x^a, (\sqrt{-|g|} \hbar \sqrt{\beta})\, \dot{x}^b).
\eea 
In this regard, the resulting metric is homogeneous of degree $0$ in $\dot{x}$ \cite{Yano1969}. Then the local coordinates on $TM_8$ is reexpressed as
\bea
x^a &=& \left(x^{a}, (\sqrt{-|g|} \hbar \sqrt{\beta}) \dot{x}^{a}\right).
\eea

The line element on tangent manifold is given as \cite{Brandt:1991hy,Brandt:1991sw},
\bea
d\tilde{s}^2 &=& g_{ab}\, dx^{a}\, dx^{b}. \label{eq:L2NewCoord}
\eea
The line element on $M$ is expressed as follows:
\bea
d\tilde{s}^2 &=& \tilde{g}_{\mu \nu}\, d\zeta^{\mu}\, d\zeta^{\nu}. \label{eq:L2OldCoord}
\eea
If the line elements in both manifolds are presumed identical, then equating Eq. \eqref{eq:L2NewCoord} and \eqref{eq:L2OldCoord} can be solved for the quantum-deformed metric \cite{Tawfik:2024gxc}
\bea
\tilde{g}_{ab}(x) &=& \left[1 + \left(-|g|\, \hbar^2\, \beta\right) |\ddot{x}|^2 \right]\,  g_{\mu\nu} = \left[1 + \mathscr{T} |\ddot{x}|^2 \right]  g_{\mu\nu}. \label{eq:dmunucurvedtilda2} 
\eea
where  $a, b$ $\in\{0,1,\cdots, 7\}$, $\alpha, \beta, \mu, \nu \in\{0,1,2,3\}$, $\ddot{x}^\lambda = \partial \dot{x}^\lambda/\partial \zeta^{\lambda}$ and $|\ddot{x}|^2\equiv\ddot{x}^\lambda \ddot{x}_\lambda = g_{\delta \gamma} \ddot{x}^{\delta} \ddot{x}^{\gamma}$ with $\lambda$, $\delta$, $\gamma$ are dummy indices.

In Section \ref{sec:tmunu1}, the derivation of the stress-energy tensor induced by quantum effects via the quantum-deformed metric is examined.

\subsection{Quantum-Induced Stress-Energy Tensor}
\label{sec:tmunu1}

Now by replacing the ordinary $\delta$ with the covariant derivatives $\nabla$ and $\eta_{\mu \nu}$ with $g_{\mu \nu}$, the action of GR in curved spacetime consists of Einstein-Hilbert action and non-gravitational part of the Lagrangian density ${\mathcal L}_{\mathtt{matter}}$, the matter field,
\bea
S &=& \int \frac{c^4}{16 \pi G} \left(R + \mathcal{L}_{\mathtt{matter}}\right) \sqrt{-g}\, d^4 x,
\eea
where $R$ is the Ricci scalar. The Jacobian term $\sqrt{-g}=\sqrt{-\mathtt{det}(g_{\mu \nu})}$, derived in section ef{sec:rgup}, guarantees  invariance of the action under diffeomorphisms, ensuring that the Lagrangian density remains scalar. The physical quantities including the Hilbert stress-energy tensor can be retrieved from demanding vanishing variation of this action with respect to the inverse metric tensor $g^{\mu \nu}$ \cite{Misner:1973prb}
\bea
T_{\mu \nu} &=& \frac{-2}{\sqrt{-|g|}}\frac{\partial}{\partial g^{\mu \nu}} \left(\sqrt{-|g|} \mathcal{L}_{\mathtt{matter}} \right) = -2 \frac{\partial \mathcal{L}_{\mathtt{matter}}}{\partial g^{\mu \nu}} + g_{\mu \nu}\, \mathcal{L}_{\mathtt{matter}} \label{eq:Tmunu2}
\eea
where $|g|=|\mathtt{det}(g_{\mu \nu})|$. It is obvious that $T_{\mu \nu}$ is merely characterized by the fundamental metric and the matter field. With the quantum-induced fundamental metric, Eq. (\ref{eq:dmunucurvedtilda2}), the stress-energy tensor, Eq. (\ref{eq:Tmunu2}), can be rewritten as
\bea
\tilde{T}_{\mu \nu} &=& -2 \frac{\partial \tilde{\mathcal{L}}_{\mathtt{matter}}}{\partial \tilde{g}^{\mu \nu}} + \tilde{g}_{\mu \nu}(x) \tilde{\mathcal{L}}_{\mathtt{matter}}. \label{eq:Tmunu4}
\eea

The quantum-induced revision of the matter Lagrangian shall be derived in section \ref{sec:qLagr}. For now, we start with quantum-deformed inverse metric
\bea
\delta \tilde{g}^{\mu \nu} = \left(1+ \mathscr{T} |\ddot{x}|^2\right)^{-1} \delta g^{\mu \nu} - \left(1+\mathscr{T} |\ddot{x}|^2 \right)^{-2} g^{\mu \nu} \mathscr{T}  \delta \left(g^{\mu \nu} \ddot{x}_\mu \ddot{x}_\nu\right). \nn
\eea
For $\tilde{g}^{\mu \nu} = (1+\mathscr{T} |\ddot{x}|^2)^{-1} g^{\mu \nu}$ and $|\ddot{x}|^2=g^{\mu \nu}\ddot{x}_{\mu} \ddot{x}_{\nu}$, one finds that
\bea
\delta \tilde{g}^{\mu \nu} &=& \left(1+\mathscr{T} |\ddot{x}|^2\right)^{-1} \delta g^{\mu \nu} -  \left(1+ \mathscr{T} |\ddot{x}|^2\right)^{-2} \mathscr{T}  \left[|\ddot{x}|^2 \delta g^{\mu \nu} + g^{\mu \nu} \ddot{x}^{\mu} \delta \ddot{x}_{\mu} + g^{\mu \nu} \ddot{x}^{\nu} \delta \ddot{x}_{\nu}\right], \label{eq:Tmunu7}
\eea
where $\ddot{x}^{\mu}=g^{\mu \nu}\ddot{x}_{\nu}$, $\ddot{x}^{\nu}=g^{\mu \nu}\ddot{x}_{\mu}$. Substituting Eq. (\ref{eq:Tmunu7}) into Eq. (\ref{eq:Tmunu4}) leads to 
\bea
\tilde{T}_{\mu \nu} &=&  \frac{-2 \delta \tilde{\mathcal{L}}_M}{\left(1+\mathscr{T} |\ddot{x}|^2\right)^{-1} \delta g^{\mu \nu} - \left(1+ \mathscr{T} |\ddot{x}|^2\right)^{-2} \mathscr{T} \left[|\ddot{x}|^2 \delta g^{\mu \nu} + g^{\mu \nu} \ddot{x}^{\mu} \delta \ddot{x}_{\mu} + g^{\mu \nu} \ddot{x}^\nu \delta \ddot{x}_\nu\right]} 
+ \tilde{g}_{\mu \nu}(x) \tilde{\mathcal{L}}_M. \label{eq:Tmunu8} \hspace*{5mm}
\eea
When replacing the small variation by differentiation and $\tilde{g}_{\mu \nu}(x)$ by $ g_{\mu \nu}$, then, the quantum-induced  stress-energy tensor can be expressed as
\bea
\tilde{T}_{\mu \nu} &=& \frac{-2\left(1+\mathscr{T} |\ddot{x}|^2\right)^2}{1 - \mathscr{T}  g^{\mu \nu} \left( \ddot{x}^{\mu} \frac{\partial \ddot{x}_{\mu}}{\partial g^{\mu \nu}} + \ddot{x}^{\nu} \frac{\partial \ddot{x}_{\nu}}{\partial g^{\mu \nu}}\right)} \frac{\partial \tilde{\mathcal{L}}_M}{\partial g^{\mu \nu}} + \left(1+\mathscr{T} |\ddot{x}|^2\right) g_{\mu \nu} \tilde{\mathcal{L}}_M. \label{eq:Tmunu12}
\eea

By comparing Eq. (\ref{eq:Tmunu12}) and Eq. (\ref{eq:Tmunu4}), one concludes that  both common terms, Eq. (\ref{eq:Tmunu7}), are entirely retrieved and multiplied by coefficients differ from unity. These coefficients depend on quantum quantities, including the unquantized metric and the variation of the second-order derivatives of tangent covectors $|\ddot{x}|^2$ with respect to the unquantized metric. As indicated by the geometric quantization, finite values of $\beta_0$ and/or $|\ddot{x}|^2$ linearly factor Eq. \eqref{eq:Tmunu12}). Upon eliminating this from the equation, the classical stress-energy tensor, represented by Eq. \eqref{eq:Tmunu4}), is evidently fully obtained. Therefore, we conclude that the stress-energy tensor given in Eq. \eqref{eq:Tmunu12}) is valid in both classical and quantum settings. We also determine that by switching $\beta_0$ (and/or $|\ddot{x}|^2$) on or off, the classical or quantum stress-energy tensor is entirely formulated. This linear factorization implies a generalization of the stress-energy tensor, owing to the complementary term that has been added to the fundamental tensor, as expressed in Eq. \eqref{eq:dmunucurvedtilda2}. To ensure thoroughness, we emphasize that the quantum revision in the stress-energy tensor is seemingly associated with appropriate changes in geometry. Further information regarding the quantum-deformed geometry of the Einstein field equation can be recalled from refs. \cite{Tawfik:2024itt} and \cite{Farouk:2023hmi}.

In Eq. \eqref{eq:Tmunu12}, the Lagrangian density for matter induced by quantum effects has yet to be derived. This topic is addressed in Section \ref{sec:qLagr}.

\subsection{Quantum-Induced Matter Lagrangian Density}
\label{sec:qLagr}

The generalized Lagrangian density can be expressed as \cite{marsden2013introduction},
\bea
\mathcal{L}(q^i,\dot{q}^i)=\frac{1}{2}\sum^n_{i,j=1} g_{i j}(q)\dot{q}^i\dot{q}^j + V(q^i), \label{eq:mmL0}
\eea
where $q^i$ are the generalized coordinates on manifold $M_4$ and $\dot{q}^i$ represent  phase-space velocities on tangent bundle $TM_4$. For any kind of matter field $\psi(x)$ regardless of its rank - be it scalar, vector, tensor, or others - where $\partial_\lambda \psi(x)$ denotes the first partial derivative of the fields and $\partial_\lambda g_{\mu \nu}(x)$ represents the first partial derivative of the metric tensor $g_{\mu \nu}$, the matter Lagrangian density is subsequently expressed as  \cite{blau2020lecture} 
\bea
\mathcal{L}_{\mathtt{matter}} (\psi(x), \partial_\lambda \psi(x), \cdots; g_{\mu \nu}(x), \partial_\lambda g_{\mu \nu}(x), \cdots). \label{eq:mmL1}
\eea
It is apparent that a quantum-induced revision of $\mathcal{L}_{\mathtt{matter}}$ would be possible, when converting $g_{\mu \nu}$ into $\tilde{g}_{\mu \nu}(x)$, Eq. (\ref{eq:dmunucurvedtilda2}), which in turn is based the minimal length uncertainty, Eq. (\ref{eq:deltaxmin2}), and the possible quantization of the spacetime,
\bea
\tilde{\mathcal{L}}_{\mathtt{matter}} (\psi(x), \partial_\lambda \psi(x), \cdots; \tilde{g}_{\mu \nu}(x), \partial_\lambda \tilde{g}_{\mu \nu}(x), \cdots). 
\label{eq:mmL2}
\eea
In this regard, we focus on two physical systems in curved spacetime; electromagnetic and Klein--Gordon (KG) field. 
\begin{itemize} 
\item Electromagnetic field  in curved spacetime is given as \cite{ydri2017lectures} 
\bea
\mathcal{L}_{\mathtt{EM}} &=& -\frac{1}{4}g^{\mu \nu}g^{\alpha \beta} F_{\mu \nu} F_{\alpha \beta}, \label{eq:mmL3}
\eea
where $F_{\mu \nu}=\partial_{\mu} A_{\nu} - \partial_{\nu} A_{\mu}$ is Faraday tensor. It is apparent that $\partial_{\mu}$ represents four-gradient and $A_{\mu}$ is four-potential. For quantum-deformed metric, Eq. (\ref{eq:dmunucurvedtilda2}), the Lagrangian density in Eq. (\ref{eq:mmL3}) can be rewritten as
\bea
\tilde{\mathcal{L}}_{\mathtt{EM}} &=& -\frac{1}{4}\tilde{g}^{\mu \nu}\tilde{g}^{\alpha \beta} F_{\mu \nu} F_{\alpha \beta} = -\frac{1}{4} \left(1+{\mathscr{T}} |\ddot{x}|^2 \right)^{-2} g^{\mu \nu}g^{\alpha \beta} F_{\mu \nu} F_{\alpha \beta} = \left(1+{\mathscr{T}} |\ddot{x}|^2\right)^{-2} \mathcal{L}_{\mathtt{EM}}. \label{eq:mmL6}
\eea
Also here, we find that for vanishing $\beta_0$ and/or $|\ddot{x}|^2$, unquantized $\mathcal{L}_{\mathtt{EM}}$ can be straightforwardly fully retrieved.
\item KG scalar field is expressed as follows \cite{padmanabhan2010gravitation}:
\bea
\mathcal{L}_{\phi} &=& - \frac{1}{2}g^{\mu \nu}\nabla_\mu \phi \nabla_\nu \phi - V(\phi), \label{eq:mmL7}
\eea
where $\phi$ is the scalar field and $V(\phi)$ is the potential field. In in curved spacetime, i.e., converting $g_{\mu \nu}$ into $\tilde{g}_{\mu \nu}(x)$, we get the quantum-induced revision version of the KG Lagrangian density,
\bea
\tilde{\mathcal{L}}_{\phi} &=& -\frac{1}{2}\tilde{g}^{\mu \nu}\nabla_\mu \phi \nabla_\nu \phi - V(\phi) = -\frac{1}{2} \left(1+ \mathscr{T} |\ddot{x}|^2 \right)^{-1}g^{\mu \nu}\nabla_\mu \phi \nabla_\nu \phi - V(\phi). \label{eq:mmL9}
\eea
By substituting $-\frac{1}{2}g^{\mu \nu}\nabla_\mu \phi \nabla_\nu \phi$ from Eq. (\ref{eq:mmL7}), the quantized scalar Lagrangian density reads 
\bea
\tilde{\mathcal{L}}_{\phi} &=&  \frac{1}{1+\mathscr{T} |\ddot{x}|^2} \left[\mathcal{L}_{\phi} - \mathscr{T} |\ddot{x}|^2 V(\phi)\right]. \label{eq:mmL10}
\eea
For vanishing $\beta_0$ and/or $|\ddot{x}|^2$, the unquantized scalar Lagrangian density ${\cal L}_{\phi}$ is fully and straightforwardly acquired.
\end{itemize}

Having determined the quantum-induced Lagrangian density for matter, the construction of the quantum-deformed stress-energy tensor, which includes the electromagnetic Lagrangian density, is elaborated in Section \ref{sec:qLagrEM}.

\subsection{Quantum-Induced Stress-Energy Tensor with EM Lagrangian Density}
\label{sec:qLagrEM}

For now, Eq. (\ref{eq:mmL6}) could be substituted into Eq. (\ref{eq:Tmunu12}). Then, the quantum-induced stress-energy tensor with EM Lagrangian density reads
\bea
\tilde{T}_{\mu \nu} &=& \frac{-2}{1 - \mathscr{T}  g^{\mu \nu} \left( \ddot{x}^{\mu} \frac{\partial \ddot{x}_{\mu}}{\partial g^{\mu \nu}} + \ddot{x}^\nu \frac{\partial \ddot{x}_\nu}{\partial g^{\mu \nu}}\right)}\frac{\partial \mathcal{L}_{\mathtt{EM}} }{\partial g^{\mu \nu}} \nn \\
&+&\frac{4 \mathscr{T} \left( \ddot{x}_\mu \ddot{x}_\nu + \ddot{x}^\mu \frac{\partial \ddot{x}_\mu}{\partial g^{\mu \nu}} + \ddot{x}^\nu\frac{\partial \ddot{x}_\nu}{\partial g^{\mu \nu}} \right)}{\left(1+\mathscr{T} |\ddot{x}|^2\right) \left[1 - \mathscr{T}  g^{\mu \nu} \left( \ddot{x}^{\mu} \frac{\partial \ddot{x}_{\mu}}{\partial g^{\mu \nu}} + \ddot{x}^\nu \frac{\partial \ddot{x}_\nu}{\partial g^{\mu \nu}}\right) \right]} \mathcal{L}_{\mathtt{EM}} \nn \\
&+& \frac{g_{\mu \nu}}{\left(1+\mathscr{T} |\ddot{x}|^2 \right)}\mathcal{L}_{\mathtt{EM}}. \label{msmtcL8} 
\eea
By substituting Eq. (\ref{eq:Tmunu2}) into Eq. (\ref{msmtcL8}), we get 
\bea
\tilde{T}_{\mu \nu} &=& \frac{1}{1 - \mathscr{T}  g^{\mu \nu} \left( \ddot{x}^{\mu} \frac{\partial \ddot{x}_{\mu}}{\partial g^{\mu \nu}} + \ddot{x}^\nu \frac{\partial \ddot{x}_\nu}{\partial g^{\mu \nu}}\right)}\left( T_{\mu \nu} - g_{\mu \nu} \mathcal{L}_{\mathtt{EM}} \right) \nn \\ 
&+&\frac{4 \mathscr{T}  \left( \ddot{x}_\mu \ddot{x}_\nu + \ddot{x}^\mu \frac{\partial \ddot{x}_\mu}{\partial g^{\mu \nu}} + \ddot{x}^\nu\frac{\partial \ddot{x}_\nu}{\partial g^{\mu \nu}}  \right)}{\left(1+\mathscr{T} |\ddot{x}|^2\right) \left[1 - \mathscr{T}  g^{\mu \nu} \left( \ddot{x}^{\mu} \frac{\partial \ddot{x}_{\mu}}{\partial g^{\mu \nu}} + \ddot{x}^\nu \frac{\partial \ddot{x}_\nu}{\partial g^{\mu \nu}}\right) \right]} \mathcal{L}_{\mathtt{EM}} \nn \\
&+& \frac{g_{\mu \nu}}{\left(1+\mathscr{T} |\ddot{x}|^2 \right)}\mathcal{L}_{\mathtt{EM}}. \label{msmtcL9} 
\eea
Furthermore, for $\partial \mathcal{L}_{\mathtt{EM}}/\partial g^{\mu \nu}= -\frac{1}{2} \left( T_{\mu \nu} - g_{\mu \nu} \mathcal{L}_{\mathtt{EM}} \right)$, Eq. (\ref{msmtcL9}) can be rewritten as 
\bea
\tilde{T}_{\mu \nu} &=& \left[1 - \mathscr{T}  g^{\mu \nu} \left( \ddot{x}^{\mu} \frac{\partial \ddot{x}_{\mu}}{\partial g^{\mu \nu}} + \ddot{x}^\nu \frac{\partial \ddot{x}_\nu}{\partial g^{\mu \nu}}\right) \right]^{-1}\, T_{\mu \nu} \nn \\
&-& \left[1 - \mathscr{T} g^{\mu \nu} \left( \ddot{x}^{\mu} \frac{\partial \ddot{x}_{\mu}}{\partial g^{\mu \nu}} + \ddot{x}^\nu \frac{\partial \ddot{x}_\nu}{\partial g^{\mu \nu}}\right)\right]^{-1}\, g_{\mu \nu}\, \mathcal{L}_{\mathtt{EM}} \nn \\
&+& \frac{4 \mathscr{T} \left( \ddot{x}_\mu \ddot{x}_\nu + \ddot{x}^\mu \frac{\partial \ddot{x}_\mu}{\partial g^{\mu \nu}} + \ddot{x}^\nu\frac{\partial \ddot{x}_\nu}{\partial g^{\mu \nu}}  \right)}{\left(1+\mathscr{T} |\ddot{x}|^2 \right) \left[1 - \mathscr{T} g^{\mu \nu} \left(\ddot{x}^{\mu} \frac{\partial \ddot{x}_{\mu}}{\partial g^{\mu \nu}} + \ddot{x}^\nu \frac{\partial \ddot{x}_\nu}{\partial g^{\mu \nu}}\right) \right]} \mathcal{L}_{\mathtt{EM}} \nn \\
&+& \frac{g_{\mu \nu}}{\left(1+\mathscr{T} |\ddot{x}|^2 \right)} \mathcal{L}_{\mathtt{EM}}. \label{msmtcL10} 
\eea
It is noteworthy to highlight that the first line of Eq. (\ref{msmtcL10}) represents the undeformed stress-energy tensor $T_{\mu \nu}$ albeit multiplied by a coefficient that is contingent upon quantum mechanical and gravitational factors, which encompass the second-order derivatives of the tangent covectors $\ddot{x}$ as well as the derivatives of $\ddot{x}$  in relation to $g_{\mu \nu}$. 

The second, third, and fourth lines express the electromagnetic Lagrangian density in a curved spacetime, multiplied by different coefficients that also rely on $\ddot{x}$ and its derivatives with respect to $g_{\mu\nu}$. Additionally, when $\beta_0$ and/or $|\ddot{x}|^2$ vanish, all contributions from $\mathcal{L}_{\mathtt{EM}}$ highlighted in the second, third, and fourth lines cease to exist because their coefficients turn to vanish as well. Also, at vanishing $\beta_0$ and/or $|\ddot{x}|^2$, the coefficient of $T_{\mu \nu}$ in the first line becomes unity, so that $\tilde{T}_{\mu \nu}=T_{\mu \nu}$ is fully obtained. 

Furthermore and by substituting Eq. (\ref{eq:mmL6}) into Eq. (\ref{msmtcL10}), $\tilde{T}_{\mu \nu}$ can be simplified to
\bea
\tilde{T}_{\mu \nu} 
&=& \left[1 - \mathscr{T} g^{\mu \nu} \left( \ddot{x}^{\mu} \frac{\partial \ddot{x}_{\mu}}{\partial g^{\mu \nu}} + \ddot{x}^\nu \frac{\partial \ddot{x}_\nu}{\partial g^{\mu \nu}}\right)\right]^{-1} \nn \\
&& \left\{T_{\mu \nu} - \left(1+\mathscr{T} |\ddot{x}|^2 \right) \mathscr{T} \left[ |\ddot{x}|^2\, g_{\mu \nu}  - 4  \frac{|\ddot{x}|^2}{g_{\mu \nu}}  - 
3  \left( \ddot{x}^{\mu} \frac{\partial \ddot{x}_{\mu}}{\partial g^{\mu \nu}} + \ddot{x}^\nu \frac{\partial \ddot{x}_\nu}{\partial g^{\mu \nu}}\right)
\right] \mathcal{L}_{\mathtt{EM}} \right\},  \label{msmtcL11} 
\eea 
from which we conclude that the quantum-deformed stress-energy tensor, associated with the electromagnetic Lagrangian density, is attained through (i) a linear factorization to $T_{\mu \nu}$ itself and (ii) a concurrent emergence of contributions from the $\mathcal{L}_{\mathtt{EM}}$-contributions. The latter are also linearly factorized with quantities depending on $\beta_0$, $|\ddot{x}|^2$, $\mathscr{T} |\ddot{x}|^2$, $g_{\mu \nu}$, and variations of the second-order derivatives of tangent covectors with respect to $g^{\mu \nu}$. Whether the $\mathcal{L}_{\mathtt{EM}}$-contributions are subtracted or added to $T_{\mu \nu}$ depends on the coefficients inside the squared brackets in front of $\mathcal{L}_{\mathtt{EM}}$. To remain within the scope of the present manuscript, the nature and significance of all these quantum-induced corrections could be studied elsewhere.

The properties and covariant derivatives of quantum-induced stress-energy tensor with EM Lagrangian are discussed in Sections \ref{sec:symmTEM} and \ref{sec:cdTEM}, respectively.

\subsubsection{Symmetric Property of Quantum-Induced Stress-Energy Tensor with EM Lagrangian}
\label{sec:symmTEM}

With the symmetry of $\mathcal{L}_{\mathtt{EM}}$ and by exchanging the covariant indexes $\mu$ and $\nu$ of Eq. (\ref{msmtcL11}), we find that
\bea
\tilde{T}_{\nu \mu} &=& \left[1 - \mathscr{T}  g^{\nu \mu} \left( \ddot{x}^{\nu} \frac{\partial \ddot{x}_{\nu}}{\partial g^{\nu \mu}} + \ddot{x}^\mu \frac{\partial \ddot{x}_\mu}{\partial g^{\nu \mu}}\right)\right]^{-1} \nn \\
&& \left\{T_{\nu \mu} - \left(1+\mathscr{T} |\ddot{x}|^2 \right)\left[g_{\nu \mu} \mathscr{T} |\ddot{x}|^2  - 4 \mathscr{T} |\ddot{x}|^2 \frac{1}{g_{\nu \mu}}  - 
3 \mathscr{T}  
 \left(\ddot{x}^{\nu} \frac{\partial \ddot{x}_{\nu}}{\partial g^{\nu \mu}} + \ddot{x}^\mu \frac{\partial \ddot{x}_\mu}{\partial g^{\nu \mu}}\right) \right] \mathcal{L}_{\mathtt{EM}} \right\}.
\label{eq:set27}
\eea
From the symmetry of metric tensor $g_{\mu \nu}=g_{\nu \mu}$ and its inverse $g^{\mu \nu}=g^{\nu \mu}$, Eq. \eqref{eq:set27} can be rewritten as
\bea
\tilde{T}_{\nu \mu} &=&
\left[1 - \mathscr{T}  g^{\mu \nu} \left( \ddot{x}^{\mu} \frac{\partial \ddot{x}_{\mu}}{\partial g^{\mu \nu}} + \ddot{x}^\nu \frac{\partial \ddot{x}_\nu}{\partial g^{\mu \nu}}\right)\right]^{-1} \nn \\
&& \left\{T_{\mu \nu} - \left(1+\mathscr{T} |\ddot{x}|^2 \right)\left[g_{\mu \nu} \mathscr{T} |\ddot{x}|^2 - 4 \mathscr{T} |\ddot{x}|^2 \frac{1}{g_{\mu \nu}}   - 
3 \mathscr{T} 
 \left( \ddot{x}^{\mu} \frac{\partial \ddot{x}_{\mu}}{\partial g^{\mu \nu}} + \ddot{x}^\nu \frac{\partial \ddot{x}_\nu}{\partial g^{\mu \nu}}\right)
\right] \mathcal{L}_{\mathtt{EM}} \right\} 
= \tilde{T}_{\mu \nu}. \hspace*{7mm} \label{eq:set29}
\eea
Thus, we conclude that similar to the unquantized stress-energy tensor, Eq. \eqref{eq:Tmunu4} \cite{carroll2019spacetime} as well as its quantum-induced version outlined, Eq. (\ref{msmtcL11}), are symmetric.

\subsubsection{Covariant Derivative of Quantum-Induced Stress-Energy Tensor with EM Lagrangian}
\label{sec:cdTEM}

Since the covariant derivative of a tensor remains consistent across all frames, we will, for the sake of simplicity, derive the covariant derivative of the stress-energy tensor with EM Lagrangian density in a freely falling frame. As the Faraday tensor exhibits antisymmetry, meaning $F_{\mu \nu}=-F_{\nu \mu}$, the covariant and partial derivatives are the same. Thus, Eq. (\ref{msmtcL10}) can be expressed as 
\bea
\tilde{T}_{\mu \nu} &= T_{\mu \nu} - \eta_{\mu \nu} \mathcal{L}_{EM} + \frac{4  \mathscr{T}  \ddot{x}_\mu \ddot{x}_\nu + \eta_{\mu \nu}}{\left(1+ \mathscr{T} |\ddot{x}|^2 \right)} \mathcal{L}_{EM}. \label{eq:cdtmumu1}
\eea
With $\partial \ddot{x}_{\mu}/\partial \eta^{\mu \nu} = \partial \ddot{x}_{\nu}/\partial \eta^{\mu \nu}=0$ and by converting the flat $\eta_{\mu \nu}$ to the curved metric tensor $g_{\mu \nu}$, then the covariant derivative reads
\bea
\nabla^\mu \tilde{T}_{\mu \nu} &=& -\nabla^\mu (\eta_{\mu \nu}\mathcal{L}_{EM}) + \nabla^{\mu} \left[\frac{4  \mathscr{T} \ddot{x}_\mu \ddot{x}_\nu + \eta_{\mu \nu}}{\left(1 + \mathscr{T} |\ddot{x}|^2 \right)} \mathcal{L}_{EM} \right]. \label{eq:cdtmumu3} 
\eea
In the preceding expression, locally vanishing $\nabla^{\mu} T_{\mu \nu}$ is substituted. Then, Eq. (\ref{eq:cdtmumu3}) leads to
\bea
\nabla^{\mu} \tilde{T}_{\mu \nu} &=& -\eta_{\mu \nu}\nabla^{\mu} \mathcal{L}_{EM} +\left[ \frac{4 \mathscr{T}  \ddot{x}_{\mu} \ddot{x}_{\nu} + \eta_{\mu \nu}}{\left(1+\mathscr{T} |\ddot{x}|^2\right)}\right]\nabla^{\mu}\mathcal{L}_{EM} \nn \\
&+& \left[\frac{\left(1+\mathscr{T} |\ddot{x}|^2\right)\left(4 \mathscr{T}  \ddot{x}_\nu\nabla^{\mu} \ddot{x}_{\mu} + 4\mathscr{T}  \ddot{x}_{\mu} \nabla^{\mu} \ddot{x}_\nu\right)}{\left(1+\mathscr{T} |\ddot{x}|^2\right)^2} \right] \mathcal{L}_{EM}, \hspace*{7mm} \label{eq:cdtmumu6}
\eea
in which $(|\ddot{x}|^2)_{,\mu}=\eta_{\mu \nu,\mu}\ddot{x}^\mu \ddot{x}^\nu=0$, $\nabla^\mu \eta_{\mu \nu}=\eta_{\mu \nu,\mu}=0$, $\nabla^\mu \ddot{x}_\mu = \ddot{x}_{\mu,\mu}=0$, and  $\nabla^\mu \ddot{x}_\nu$ are utilized.  Then, 
\bea
\nabla^{\mu} \tilde{T}_{\mu \nu} &=& \left\{\left[\frac{4 \mathscr{T} |\ddot{x}|^2 + \eta_{\mu \nu}}{\left(1+\mathscr{T} |\ddot{x}|^2\right)}\right] - \eta_{\mu \nu}\right\} \nabla^{\mu} \mathcal{L}_{EM}. \label{eq:cdtmumu7}
\eea
Thus, we conclude that $\nabla^{\mu} \tilde{T}_{\mu \nu}$ results in the covariant derivative of the Lagrangian density, multiplied by coefficients that are contingent upon the metric tensor, $\beta_0$, $|\ddot{x}|^2$, and $\mathscr{T}$. If the quantization is eliminated by making $\beta_0$ and/or $|\ddot{x}|^2$ equal to zero, $\nabla^{\mu} \tilde{T}_{\mu \nu}$ will also vanish, even when $\nabla^{\mu} \mathcal{L}_{EM}$ is finite. The latter is defined in Eq. (\ref{eq:mmL3}) in free falling frame, for which $g$ is to be replaced by $\eta$,
\bea
\nabla^{\mu} \mathcal{L}_{EM} &=& -\frac{1}{4} \eta^{\mu \nu} \eta^{\alpha \beta} \left(F_{\alpha \beta} F_{\mu \nu ,\mu} + F_{\mu \nu} F_{\alpha \beta ,\mu} \right). \label{eq:cdtmumu9}
\eea
Maxwell theory implies $\partial_{\mu} F^{\mu \nu}=\mu_0 J^{\nu}$ with $\mu_0$ is permeability of free space and $J^{\nu}$ is four-current. From Eq. (\ref{eq:cdtmumu7}) and Eq. (\ref{eq:cdtmumu9}), we get
\bea
\nabla^\mu \tilde{T}_{\mu \nu} &=& \left\{\left[\frac{4 \mathscr{T} |\ddot{x}|^2\eta_{\mu \nu} + \eta_{\mu \nu}}{\left(1+\mathscr{T} |\ddot{x}|^2 \right)}\right] - \eta_{\mu \nu}\right\} \left[-\frac{1}{4} \eta^{\mu \nu} \eta^{\alpha \beta} \left(F_{\alpha \beta} F_{\mu \nu ,\mu} + F_{\mu \nu} F_{\alpha \beta ,\mu} \right)\right], \label{eq:cdtmumu10}
\eea
where $F_{\mu \nu , \mu} = \left(\eta_{\mu \alpha} F^{\alpha \beta} \eta_{\nu \beta} \right)_{, \mu}$ and $F_{\alpha \beta , \mu} = \left(\eta_{\alpha \mu} F^{\mu \nu} \eta_{\beta \nu} \right)_{, \mu}$.

We conclude that
\begin{itemize} 
\item for unquantized stress-energy tensor, i.e., vanishing $\beta_0$ and/or $|\ddot{x}|^2$, then
\bea
\nabla^\mu \tilde{T}_{\mu \nu} &=& 0, \label{eq:cdtmumu11}
\eea
\item for quantum-induced  stress-energy tensor, i.e., finite $\beta_0$ and/or $|\ddot{x}|^2$, in vacuum spacetime, i.e., vanishing four-current, we also get
\bea
\nabla^\mu \tilde{T}_{\mu \nu} &=& 0, \label{eq:cdtmumu11b}
\eea
as $F_{\mu \nu ,\mu}=F_{\alpha \beta ,\mu}=0$. 
\end{itemize}
Otherwise, we observe that $\nabla^\mu \tilde{T}_{\mu \nu}$ exhibits divergence. In such a scenario, the Lagrangian density ought to be associated with a charged particle serving as a source in the inhomogeneous Maxwell equations so that
\bea
\nabla^\mu (\tilde{T}^{\mathtt{EM}}_{\mu \nu}+\tilde{T}^{\mathtt{particle}}_{\mu \nu}) &=& 0, \label{eq:cdtmumu12}
\eea

Let us now revisit the notion that these results appear to correspond with the logical principles of the Noether theorem regarding translation
\cite{Noether1918}; every symmetry found in Nature is associated with a conservation law. The null covariant derivative of the stress-energy tensor arises from SR, specifically, i.e., $\delta^{\mu}~T_{\mu \nu}=0$. Although the Noether theorem does not serve as the primary justification for extending this result to quantum-modified GR, it is fundamentally rooted in the comprehension that GR and SR are locally equivalent, such that in local frames, for example, a free-falling frame, $\delta^{\mu}T_{\mu \nu}=\nabla^{\mu}T_{\mu \nu}=0$. For now, we can also interpret vanishing $\nabla^\mu \tilde{T}_{\mu \nu}$ due to diffeomorphism invariance. As long as the metric tensor could not be associated with a Killing vector-field, no conserved quantity could be associated with $\nabla^{\mu}\tilde{T}_{\mu \nu}=\nabla^{\mu}T_{\mu \nu}=0$ \cite{carroll2019spacetime}.

In Section \ref{sec:Bianchi}, the  contracted Bianchi identity and conservation are examined to demonstrate the vanishing covariant divergence of the Einstein tensor, while highlighting that this does not generally apply to the complete left-hand side of a modified gravitational equation. For the comprehensive generalized equations, we explicitly verify that the covariant divergence of the left-hand side corresponds with the outcome for the covariant divergence of the energy-stress tensor, regardless of whether the result is vanishing or non-vanishing.

\subsection{Contracted Bianchi Identity and Conservation}
\label{sec:Bianchi}

From the geometric quantization approach, the schematic formulation of EFE can be expressed as \cite{Tawfik:2025wsl,Tawfik:2025ldp,Tawfik:2024itt,Tawfik:2024bdf,NasserTawfik:2024afw,Farouk:2023hmi,Tawfik:2023ugm,Tawfik:2023hdi,Tawfik:2023orl,Tawfik:2023kxq,Tawfik:2020zvf,TawfikAN2023b,
TawfikAN2023a}
\bea
G_{\mu\nu} + \Delta_{\mu\nu} = 8 \pi G T_{\mu\nu},
\eea
where $G_{\mu\nu}$ is Einstein tensor based on conventional metric and $\Delta{\mu\nu}$ combines extra geometric terms stemming from the geometric quantization. Let us assume $E_{\mu\nu}=G_{\mu\nu} + \Delta_{\mu\nu}$. From conventional contracted Bianchi identity, i.e.,
\bea
\nabla_{\mu}\, G^{\mu\nu} = 0,
\eea
the divergence in $E_{\mu\nu}$ is given as
\bea
\nabla_{\mu}\, E_{\mu\nu} = \nabla_{\mu}\, \left(G_{\mu\nu} + \Delta_{\mu\nu}\right) = \nabla_{\mu}\, \Delta^{\mu\nu}.
\eea
Therefore, there are only two consistent options:  
\begin{enumerate}
\item Diffeomorphism‑invariant construction (most natural): 
If the quantum-deformed action remains diffeomorphism invariant and $\Delta_{\mu\nu}$ based on varying scalar action with respect to the fundamental metric, then the Noether identity leads to the geometric identity
\bea
\nabla_{\mu}\, E^{\mu\nu} = 0, \qquad \mathtt{and} \qquad \nabla_{\mu}\, \Delta^{\mu\nu}=0.
\eea
This allows for the standard covariant conservation of the possibly quantum‑corrected stress-energy tensor,
\bea
\nabla_{\mu}\, T^{\mu\nu}=0.
\eea
\item Effective‑theory viewpoint (more general):  If $\Delta_{\mu\nu}$ is not divergence‑free by construction, we find that 
\bea
\nabla_{\mu}\, T^{\mu\nu} =   \frac{1}{8 \pi G} \nabla_{\mu} E^{\mu\nu}.
\eea
This means that the left‑hand side is not divergence‑free, and the price to be payed is an effective non‑conservation of the matter stress-energy tensor that can be interpreted as exchange of energy-momentum between “quantum geometry” and matter.

The question to be answered now is what does the geometric quantization actually suggest? Let start by highlighting that the geometric quantization is (i) based on a geometric quantization of the fundamental metric or phase‑space metric and (ii) typically derived from a diffeomorphism‑invariant action or at least intended to be compatible with GR’s symmetry structure. Then, the natural and consistent reading is (i) the full geometric side, left-hand, including $\Delta_{\mu\nu}$ is constructed so that $\nabla_{\mu} E^{\mu\nu} =0$ holds identically and (ii) consequently, any quantum corrections to the stress-energy tensor for scalar and vector fields, for instance, can be arranged so that
\bea
\nabla _{\mu}\, T^{\mu\nu}_{\mathtt{total, q-corrected}} = 0
\eea  
\end{enumerate}
To summarize, if it is found - in any alternative model for GR - that $\nabla_{\mu}\, \Delta^{\mu\nu} \neq 0$, the proper interpretation would be formulating a proper description where matter is not separately conserved, because it is exchanging energy-momentum with the quantized geometric sector.

The second question is how does the quantum-modified left-hand side appear? To answer it, let us write
\bea
\langle\hat{G}_{\mu\nu}(\hat{g})\rangle = G_{\mu\nu}(g) + Q_{\mu\nu}(g, \partial g, \cdots),
\eea
where $G_{\mu\nu}(g)$ is the conventional Einstein tensor of a "background" or mean metric, and $Q_{\mu\nu}$ combines all quantum-geometric corrections emerged from the quantum-deformed fundamental metric. Now there are two choices to formulate EFE:
\begin{enumerate}
\item Collect all geometry on the left so that 
\bea
G_{\mu\nu} + Q_{\mu\nu} = 8 \pi T_{\mu\nu}^{\mathtt{classic}},
\eea
\item Keep Einstein tensor on the left, move quantum quantities to the right,
\bea
G_{\mu\nu} &=& 8 \pi G \left(T_{\mu\nu}^{\mathtt{classic}} + T_{\mu\nu}^{\mathtt{quantum}}\right),
\eea
where $ T_{\mu\nu}^{\mathtt{quantum}}=-Q_{\mu\nu}/8 \pi G$.
\end{enumerate}
In all our publications so far, the second arrangement is explicitly adopted \cite{Tawfik:2025wsl,Tawfik:2025ldp,Tawfik:2024itt,Tawfik:2024bdf,NasserTawfik:2024afw,Farouk:2023hmi,Tawfik:2023ugm,Tawfik:2023hdi,Tawfik:2023orl,Tawfik:2023kxq,Tawfik:2020zvf,TawfikAN2023b,
TawfikAN2023a}. Also, the present study explicitly adopts the quantum corrections to be interpreted as a quantum-induced stress-energy tensor. Because the Bianchi identity still holds for the Einstein tensor of the (mean) metric, namely $\nabla_{\mu} G^{\mu}_{\nu} =0$, the consistent conservation law leads to
\bea
\nabla_{\mu}\, \left(T^{\mu\; \mathtt{classic}}_{\nu} + T^{\mu\; \mathtt{quantum}}_{\nu}\right)=0
\eea
Equivalently, in the “all-geometry-on-the-left” formulation,
\bea
\nabla_{\mu}\, \left(G^{\mu}_{\nu} + Q^{\mu}_{\nu}\right) = 0 \;\;\; \Leftrightarrow \nabla_{\mu}\, T^{\mu\; \mathtt{classic}}_{\nu} = \frac{1}{8 \pi G} \nabla_{\mu} Q^{\mu}_{\nu}.
\eea
Thus, the full LHS is divergence-free, and the total stress-energy, i.e., classical and quantum-induced, is covariantly conserved. On the other hand, all quantum corrections are kept on the left and only a classical $T_{\mu\nu}$ on the right, then one generically finds that
\bea
\nabla_{\mu}\, \left(G^{\mu}_{\nu} + Q^{\mu}_{\nu}\right)\neq 0.
\eea
Thus,  the usual conservation law for the classical $T_{\mu\nu}$. Again, by reinterpreting the quantum corrections as effective stress-energy tensor, the structure of GR is kept and 
\bea
\nabla_{\mu}\, T^{\mu\; \mathtt{total}}_{\nu} = 0.
\eea

To summarize, we find that from the geometric quantization approach and if the Einstein tensor that based on the quantized (or mean) metric form the left-hand side, the covariant divergence likely vanishes by the contracted Bianchi identity. Any extra geometric terms arising from the proposed quantization must either (i) be moved to the right-hand side and treated as a quantum-induced stress-energy tensor, so that the total $T_{\mu\nu}$ is conserved, or else (ii) they make the full geometric left-hand side not divergence-free, signaling a modified conservation law.

Section \ref{sec:QMscal}, the derivation of quantum-induced stress-energy tensor with scalar Lagrangian density is presented.

\subsection{Quantum-Induced Stress-Energy Tensor with Scalar Lagrangian Density}
\label{sec:QMscal}

For spacetime filled with scalar field and by substituting Eq. (\ref{eq:mmL9}) into Eq. (\ref{eq:Tmunu12}), the quantum-induced stress-energy tensor can be expressed as
\bea
\tilde{T}_{\mu \nu} &=& \frac{-2\left(1+\mathscr{T} |\ddot{x}|^2\right)^2}{1 - \mathscr{T}  g^{\mu \nu} \left(\ddot{x}^{\mu} \frac{\partial \ddot{x}_{\mu}}{\partial g^{\mu \nu}} + \ddot{x}^\nu \frac{\partial \ddot{x}_\nu}{\partial g^{\mu \nu}}\right)}  \frac{\partial}{\partial g^{\mu \nu}} \left[-\frac{1}{2} \left(1+\mathscr{T} |\ddot{x}|^2\right)^{-1} g^{\mu \nu}\nabla_\mu \phi \nabla_\nu \phi - V(\phi) \right] \nn \\
&+& \left(1+\mathscr{T} |\ddot{x}|^2\right) g_{\mu \nu}  \left[-\frac{1}{2} \left(1+\mathscr{T} |\ddot{x}|^2\right)^{-1}g^{\mu \nu}\nabla_\mu \phi \nabla_\nu \phi - V(\phi) \right], \label{qsetKG12} 
\eea
With $\ddot{x}^\mu = g^{\mu \nu} \ddot{x}_\nu$ and $\ddot{x}^\nu = g^{\mu \nu} \ddot{x}_\mu$, we find that
%
%
%
\bea
\tilde{T}_{\mu \nu} &=& - \frac{2\left(1+\mathscr{T} |\ddot{x}|^2\right)^2}{1 - \mathscr{T}  g^{\mu \nu} \left( \ddot{x}^{\mu} \frac{\partial \ddot{x}_{\mu}}{\partial g^{\mu \nu}} + \ddot{x}^\nu \frac{\partial \ddot{x}_\nu}{\partial g^{\mu \nu}}\right)}  \frac{\partial {\cal L}_{\phi}}{\partial g^{\mu \nu}} 
+ g_{\mu \nu} {\cal L}_{\phi} \nn \\
&-& \frac{1}{2} \frac{\mathscr{T}  \left(1+\mathscr{T} |\ddot{x}|^2\right)^{-2}  |\ddot{x}|^2 
+ \mathscr{T} (1+\mathscr{T} |\ddot{x}|^2)^{-2} 
}
{1 - \mathscr{T}  g^{\mu \nu} \left(\ddot{x}^{\mu} \frac{\partial \ddot{x}_{\mu}}{\partial g^{\mu \nu}} + \ddot{x}^\nu \frac{\partial \ddot{x}_\nu}{\partial g^{\mu \nu}}\right)}  \left(\ddot{x}^\mu \frac{\partial \ddot{x}_\mu}{\partial g^{\mu \nu}} + \ddot{x}^\nu \frac{\partial \ddot{x}_\nu}{\partial g^{\mu \nu}} \right) g^{\mu \nu} \nabla_\mu \phi \nabla_\nu \phi \nn \\
&-& g_{\mu \nu} \mathscr{T} |\ddot{x}|^2 \, V(\phi). \label{qsetKG20}
\eea
For vanishing $\beta_0$ and/or $|\ddot{x}|^2$, the right-hand side of Eq. (\ref{qsetKG20}) turns to express unquantized $T_{\mu \nu}$, where the first line gets a unity factor, i.e., the entire coefficient $(1+ \mathscr{T} |\ddot{x}|^2[1 - \mathscr{T} g^{\mu \nu} (\ddot{x}^{\mu} \frac{\partial \ddot{x}_{\mu}}{\partial g^{\mu \nu}} + \ddot{x}^\nu \frac{\partial \ddot{x}_\nu}{\partial g^{\mu \nu}})]^{-1} \rightarrow 1$. Also, both second and third lines entirely vanish. The nature and significance of the additional contributions associated with the geometric quantization, namely the second and third lines, could be studied elsewhere.

The properties and Covariant derivatives of the quantum-induced  stress-energy tensor with scalar Lagrangian density are discussed in Sections \ref{sec:symmTscal} and \ref{sec:codrTEM}, respectively.

\subsubsection{Symmetric Property of Quantum-Induced Stress-Energy Tensor with Scalar Lagrangian Density}
\label{sec:symmTscal}

By exchanging the indexes $\mu$ and $\nu$, Eq. (\ref{qsetKG20}) reads
\bea
\tilde{T}_{\nu \mu} &=& - \frac{2\left(1+\mathscr{T} |\ddot{x}|^2\right)^2}{1 - \mathscr{T}  g^{\nu \mu} \left(\ddot{x}^{\nu} \frac{\partial \ddot{x}_{\nu}}{\partial g^{\nu \mu}} + \ddot{x}^\mu \frac{\partial \ddot{x}_\mu}{\partial g^{\nu \mu}}\right)}  \frac{\partial {\cal L}_{\phi}}{\partial g^{\nu \mu}} 
+ g_{\nu \mu} {\cal L}_{\phi} - g_{\nu \mu} \mathscr{T} |\ddot{x}|^2 \, V(\phi) \nn \\
&-& \frac{1}{2} 
\frac{
\mathscr{T}  \left(1+\mathscr{T} |\ddot{x}|^2\right)^{-2} |\ddot{x}|^2  
+ \mathscr{T} |\ddot{x}|^2  
\left(1+\mathscr{T} |\ddot{x}|^2\right)^{-2} 
}
{1 - \mathscr{T}  g^{\nu \mu} \left(\ddot{x}^{\nu} \frac{\partial \ddot{x}_{\nu}}{\partial g^{\nu \mu}} + \ddot{x}^{\mu} \frac{\partial \ddot{x}_\mu}{\partial g^{\nu \mu}} \right)}  \left(\ddot{x}^\nu \frac{\partial \ddot{x}_\nu}{\partial g^{\nu \mu}} + \ddot{x}^\mu \frac{\partial \ddot{x}_\mu}{\partial g^{\nu \mu}} \right) g^{\nu \mu} \nabla_\nu \phi \nabla_\mu \phi.
\eea
The symmetry of $g_{\mu \nu}=g_{\nu \mu}$ and $g^{\mu \nu}=g^{\nu \mu}$ leads to
\bea
\tilde{T}_{\nu \mu} &=& - \frac{2\left(1+\mathscr{T} |\ddot{x}|^2\right)^2}{1 - \mathscr{T}  g^{\mu \nu} \left( \ddot{x}^{\mu} \frac{\partial \ddot{x}_{\mu}}{\partial g^{\mu \nu}} + \ddot{x}^\nu \frac{\partial \ddot{x}_\nu}{\partial g^{\mu \nu}}\right)}  \frac{\partial {\cal L}_{\phi}}{\partial g^{\mu \nu}} 
+ g_{\mu \nu} {\cal L}_{\phi} - g_{\mu \nu} \mathscr{T} |\ddot{x}|^2\, V(\phi) \nn \\
&-& \frac{1}{2} \frac{\mathscr{T}  \left(1+\mathscr{T} |\ddot{x}|^2\right)^{-2}  
|\ddot{x}|^2 
+ \mathscr{T} |\ddot{x}|^2\left(1+\mathscr{T} |\ddot{x}|^2 \right)^{-2}
}
{1 - \mathscr{T} g^{\mu \nu} \left( \ddot{x}^{\mu} \frac{\partial \ddot{x}_{\mu}}{\partial g^{\mu \nu}} + \ddot{x}^\nu \frac{\partial \ddot{x}_\nu}{\partial g^{\mu \nu}}\right)} \left(\ddot{x}^\mu \frac{\partial \ddot{x}_\mu}{\partial g^{\mu \nu}} + \ddot{x}^\nu \frac{\partial \ddot{x}_\nu}{\partial g^{\mu \nu}} \right) g^{\mu \nu}\nabla_\mu \phi \nabla_\nu \phi
= \tilde{T}_{\mu \nu}. \hspace*{7mm}
\eea
Thus, we conclude that $\tilde{T}_{\mu \nu}$ with scalar Lagrangian seems to fulfill the symmetry property. 

The covariant derivative of the stress-energy tensor induced by quantum effects, which is associated with the scalar Lagrangian, is detailed in Section \ref{sec:codrTEM}.

\subsubsection{Covariant Derivative of Quantum-Induced Stress-Energy Tensor with Scalar Lagrangian}
\label{sec:codrTEM}

The Noether theorem establishes a connection between differentiable symmetries and conservation laws within the fundamental framework of a theory \cite{Noether1918}. This differentiable symmetry is associated with the action of a physical system, which defines the  behavior of the system through the principle of least action. In section \ref{sec:symmTscal}, we have illustrated that $\tilde{T}_{\mu\nu}=\tilde{T}_{\nu\mu}$. Consequently, we  categorically conclude that $\tilde{T}_{\mu\nu}$ complies with the conservation laws, locally. On one hand, the conservation of angular momentum is ensured by the symmetrical nature of the spacetime indexes of the quantum-induced stress-energy tensor \cite{bak1994energy}. On the other hand, the conservation of energy-momentum is satisfied by the vanishing covariant derivative of the stress-energy tensor with respect to spacetime. The quantum-induced revision of the stress-energy tensor discussed in this paper is defined based on the Hilbert action, $\delta S=0$. The vanishing variation of action $S$ relates to the property of diffeomorphism invariance, which implies that the stress-energy tensor is also invariant under diffeomorphisms, and its covariant derivative results
\bea
\nabla^\mu \tilde{T}_{\mu \nu}&=&0.
\eea
i.e., the stress-energy tensor is not explicitly depending on the coordinates $x^{\mu}$ or its divergence is zero (locally conserved).

The potential signals, measurements, and cosmological observations for quantum-deformed stress-energy tensor are discussed in Section \ref{sec:Sgnls}.

\subsection{Possible Signals, Measurements and Observations}
\label{sec:Sgnls}

The quantum corrections in the stress-energy tensor enhance the conventional symmetric $T_{\mu\nu}$ by incorporating additional terms that are contingent upon (i) the minimal length/RGUP scale \cite{Tawfik:2024gxc}, (ii) the derivatives of tangent covectors within the phase-space bundle, and (iii) coefficients that are dependent on the metric, which include $G$, $\hbar$, $c$, and $\ell_{\mathtt{pl}}$. If formally still have $\nabla_{\mu\nu}^{\mathtt{classic}}=0$ to that - as discussed earlier- structure of EFE is preserved, but the content of $T_{\mu\nu}$ now carries quantum-geometry corrections to energy density, pressure, and effective potentials. Both symmetry and modified expansion, along with early Universe phenomenology \cite{Chokyi:2026ryf,Tawfik:2021rvv,Tawfik:2012he,Tawfik:2011mw}, are set to undergo changes. This is due to the fact that the Einstein equations are expressed as follows: 
\bea
G_{\mu\nu} &=& 8 \pi G\, T_{\mu\nu}^{\mathtt{quantum}}=0,
\eea
and accordingly any correction to $T_{\mu\nu}$, as shown in this study, feeds directly into the Friedmann and Raychaudhuri equations.

Potential observable effects may include: (a) modifications in the effective equation of state, where quantum terms could simulate dark energy, stiff matter, or components resembling bulk viscosity, potentially influencing the expansion history $H(z)$ and the deceleration parameter, (b) the softening of singularities or their replacement with a bounce, as quantum-induced contributions to energy density and pressure may violate or modify classical energy conditions, thereby permitting nonsingular cosmologies or bounces, which can be observed through constraints on early-time behavior derived from CMB and primordial gravitational waves, and (c) variations in corrections to primordial perturbations, which differ if the quantum stress-energy modifies the dynamics of scalar and tensor perturbations, potentially leaving traces in the CMB power spectrum, spectral index, and tensor-to-scalar ratio. All these aspects could be investigated through CMB observations (Planck, LiteBIRD, CMB S4), BAO/SNe/standard sirens for $H(z)$, and the primordial gravitational wave background resulting from inflation or bounce scenarios.

Not only does the early Universe present intriguing phenomena \cite{Chokyi:2026ryf,Tawfik:2021rvv,Tawfik:2012he,Tawfik:2011mw}, but compact objects and systems with strong gravitational fields also play a significant role. The stress-energy tensor, influenced by quantum effects, modifies the matter content in EFE, even for what are considered 'classical' fields such as scalar, electromagnetic, and fluid fields. Possible signatures of these effects include: (a) Neutron stars and exotic compact objects, where a modified pressure-density relation, i.e., effective EoS, arising from quantum contributions can modify mass-radius relationships, maximum mass limits, and tidal deformabilities, which can be tested using NICER, as well as LIGO/Virgo/KAGRA constraints on tidal Love numbers; (b) The structure of black holes and their accretion processes, where quantum corrections to the stress-energy of electromagnetic and scalar fields in proximity to horizons can affect the horizon radius and the geometry near the horizon, the structure of the accretion disk, and the innermost stable circular orbit (ISCO) radius, thereby influencing X-ray spectra and iron line profiles, as well as the size and shape of black hole shadows as observed by the Event Horizon Telescope (EHT); and (c) Gravitational wave signals, where the inspiral and ringdown phases are contingent upon both the geometry and the matter content, with quantum modifications to $T_{\mu\nu}$ potentially shifting quasi-normal mode frequencies and tidal effects in binary neutron star and black hole-neutron star mergers.

When searching for coherent deviations across various sectors that all indicate the same fundamental quantum-induced $T_{\mu\nu}$: (i) A consistent alignment of cosmological data (CMB + BAO + SNe + GW standard sirens) that favors an effective fluid or early-time behavior that naturally emerges from the quantum-deformed stress-energy tensor, rather than arbitrary dark components, (b) Significant gravitational anomalies - in black hole shadows, QNM spectra, or neutron star mass-radius relationships - that correspond with the predictions of the modified $T_{\mu\nu}$ and metric collectively, (c) Planck suppressed yet nonzero signatures of minimal length in the propagation of high-energy photons/neutrinos that align with the same GUP scale incorporated into the stress-energy tensor.

Last but not least for phase space / minimal length effects: the dynamics of deformed matter arise from the construction existing within a phase space governed by RGUP \cite{Tawfik:2024gxc}. The stress-energy tensor incorporates minimal-length and noncommutative effects within matter fields. The phenomenological windows fit into three categories: (a) High energy astrophysics, where modified dispersion relations and effective potentials for photons and scalar fields may manifest as energy-dependent propagation (e.g., time of flight delays from GRBs and AGN flares) and slight deviations from Lorentz invariance at Planck suppressed levels; (b) Early universe particle production, where quantum-induced corrections to scalar and gauge field stress-energy can influence reheating, particle creation, and relic abundances; and (c) Laboratory-adjacent regimes, where in principle, RGUP-induced corrections \cite{Tawfik:2024gxc} to stress-energy tensor could relate to precision tests of gravity using quantum systems such as atom interferometry and optomechanics, although the current geometric quantization framework primarily focuses on relativistic and astrophysical contexts.

\section{Conclusions and Outlook}
\label{sec:Cncls}

Approaches like doubly-special relativity, string theory, loop quantum gravity, and noncommutative geometry, for example, introduce the discretization to spacetime. Thereby, the interactions of finite gravitational field could be integrated into QM. Consequently, the generalized noncommutative Heisenberg algebra and the generalized uncertainty principle (GUP) have been proposed to extend QM to encompass gravity and the fundamental metric. On the other hand, Finsler geometry, which serves as a generalization of Riemann geometry, represents a natural framework for examining the postulates of GR and investigating the quantization of spacetime, including its connections and curvatures. By considering the minimal length uncertainty and the additional curvature applied to the eight-dimensional tangent bundle, one could impose quantization on the four-dimensional Riemann manifold of spacetime. This approach ensures the most comprehensive geometric length measure for curves, particularly when the invariance under local Lorentz transformations is relaxed. The generalized (quantized) fundamental metric is to be determined by the Hessian matrix in the extended coordinates ($x^\mu, (\sqrt{-|g|} \hbar \sqrt{\beta_0}) \dot{x}^\mu)$ on the tangent bundle. To summarize, GR is assumed to be quantized, where principles of QM, such as noncommutation and superposition principle are retained, while GR is modified and also QM is conjectured to be gravitized, where principles of GR, such as equivalence and relativity principle are retrained, while their modifications of QM are taken into consideration. The quantum-induced version of the torsion-free fundamental tensor is constructed through a complementary term reconciling the principles of QM and GR and comprising generalized noncommutative Heisenberg algebra. In this context, any effort to reconcile the principles of GR and QM are associated with challenges faced by notable scientists over the last century. Alternative theories have not succeeded in offering satisfactory resolutions. The main obstacle resides in the approximations that are intrinsic to both theories. GR, in its conventional formulation,  emphasizes on gravity, while QM is not accounting for gravitational fields. The contributions of one of the authors (AT) arise from an initiative to mitigate these approximations and to propose a new theory regarding the kinematics of quantum particles within curved spacetime. This paper investigates the recent attempt to extend Riemann geometry to Finsler geometry and subsequently to Hamilton geometry. The generalization of QM seeks to integrate finite gravitational fields, particularly concerning the Heisenberg uncertainty principle.

The present script introduces an attempt for quantum-induced revision of the stress-energy tensor, which is solely based on quantum-induced corrections added to the fundamental metric. For matter Lagrangian density with electromagnetic and scalar fields, we have constructed the stress-energy tensor. For both fields, the quantization is achieved through (i) a linear factorization to the stress-energy tensor itself and (ii) simultaneous emergence of additional contributions of the Lagrangian densities. The latter are also linearly factorized with quantities depending on RGUP parameter $\beta_0$, the second-order derivatives of tangent covectors $|\ddot{x}|^2$, the fundamental metric $g_{\mu \nu}$, and the derivatives of $|\ddot{x}|^2$ with respect to $g^{\mu \nu}$. 

We have shown that both quantum-induced stress-energy tensors with electromagnetic and with scalar Lagrangian densities are symmetric in their covariant indexes. For the quantum-induced stress-energy tensor with EM Lagrangian density in vacuum spacetime, we have derived the covariant derivative and categorically concluded its vanishing divergence. Otherwise, the Lagrangian density should be coupled with matter source in the inhomogeneous Maxwell equation. For the quantum-induced  stress-energy tensor with scalar Lagrangian density, the conclusion of vanishing covariance derivative is based on the Noether theorem for translation that very symmetry implies conservation. 

Fortunately, the geometric quantization proposed in the present script is linearly factorized to the unquantized stress-energy tensor. The unquantized version of the stress-energy tensor is the basis and is always present. The quantized version appears as supplementary, which could be switched off and on. In both cases, the unquantized stress-energy tensor is present. On the other hand, the quantization introduces quantum coefficients and functions, while the relativistic quantities differ from unity in relation to the unquantized stress-energy tensor. Additionally, it contributes further elements from the Lagrangian densities and potentials, which include coefficients related to quantum and relativistic quantities. The nature and relevance of these coefficients, along with the extra Lagrangian densities and potentials, could be analyzed elsewhere.

\

\bibliographystyle{unsrtnat}
\bibliography{2022-07-07-QuantizationEnergyMomentumTensor}

\end{document}